# Efficiency Enhancement in Organic Solar Cells by Incorporating Silica-coated Gold Nanorods at the Buffer/Active interface

Haoyang Zhao,[a] Fan Yang,[a] Peiqian Tong,[a] Yanxia Cui,[a,b,*] Yuying Hao,[a,*] Qinjun Sun,[a] Fang Shi,[a] Qiuqiang Zhan,[c] Hua Wang,[d] and Furong Zhu[a,b,*]

Abstract: The performance of organic solar cells (OSCs) can be greatly improved by incorporating silica-coated gold nanorods (Au@SiO$_2$ NRs) at the interface between the hole transporting layer and the active layer due to the plasmonic effect. The silica shell impedes the aggregation effect of the Au NRs in ethanol solution as well as the server charge recombination on the surface of the Au NRs otherwise they would bring forward serious reduction in open circuit voltage when incorporating the Au NRs at the positions in contact with the active materials. As a result, while the high open circuit voltage being maintained, the optimized plasmonic OSCs possess an increased short circuit current, and correspondingly an elevated power conversion efficiency with the enhancement factor of ~11%. The origin of performance improvement in OSCs with the Au@SiO$_2$ NRs was analyzed systematically using morphological, electrical, optical characterizations along with theoretical simulation. It is found that the broadband enhancement in absorption, which yields the broadband enhancement in exciton generation in the active layer, is the major factor contributing to the increase in the short circuit current density. Simulation results suggest that the excitation of the transverse and longitudinal surface plasmon resonances of individual NRs as well as their mutual coupling can generate strong electric field near the vicinity of the NRs, thereby an improved exciton generation profile in the active layer. The incorporation of Au@SiO$_2$ NRs at the interface between the hole transporting layer and the active layer also improves hole extraction in the OSCs.

[a]Key Laboratory of Advanced Transducers and Intelligent Control System (Ministry of Education), Taiyuan University of Technology, Taiyuan 030024, China.
[b]Department of Physics and Institute of Advanced Materials, Hong Kong Baptist University, Hong Kong, China.
[c]Centre for Optical and Electromagnetic Research, South China Academy of Advanced Optoelectronics, South China Normal University, Guangzhou 510006, China.
[d]Key Laboratory of Interface Science and Engineering in Advanced Materials, Taiyuan University of Technology, Taiyuan 030024, China.
* Authors to whom correspondences should be addressed:yanxiacui@gmail.com, haoyuyinghyy@sina.com, frzhu@hkbu.edu.hk.

## 1. Introduction

Organic solar cells (OSCs)[1] are promising candidates in the field of photovoltaic technology owing to their material advantage as well as compatibility with roll-to-roll fabrication and industry. Recent progress of OSCs demonstrated that the power conversion efficiency (PCE) has been improved to around 12%[2], and industry investment on high throughput manufacturing of OSCs at low cost has already emerged[3]. The potential applications of OSCs include solar plants, energy supply for portable electric devices, semi-transparent solar cell windows, building or public facility applications, etc.

Due to the relatively short exciton diffusion length and low carrier mobility of organic semiconductors, the featured thickness of the active layer in OSCs is limited (~100 nm). This reduces the absorption efficiency, thereby a significant loss in photon-to-exciton generation, the limitation all types of thin-film solar cells face. In OSCs, a viable scheme is to boost light absorption in the active layer to the best extent possible. To date, different light trapping methods[4-5] have been reported for improving absorption in solar cells, including the use of anti-reflection coatings[6], photonic crystals[7], and plasmonic nanostructures[8-10]. Among them, plasmonic nanostructures are a superior approach in improving light absorption in solar cells due to the near-field enhancement caused by the localized surface plasmon (SP) resonances[11]. Plasmonic elements used for light management can be categorized into two groups:

nanogratings[12-14] and nanoparticles (NPs)[15-16]. Plasmonic nanogratings are generally in well-organized patterns, which require complicated nanofabrication techniques, bringing challenges to cost management. Plasmonic NPs, which can be synthesized by low-cost solution processes, are compatible with the fabrication of OSCs and offer a better choice of light trapping components in OSCs.

Plasmonic NPs are usually made of noble metals such as gold (Au) and silver (Ag), having dimensions much smaller than the wavelength of visible light. There are a plethora of different particles in this realm featured of diverse compositions, shapes, and sizes. Selection and the distribution of NPs within OSCs play very important roles in determining the performance of PCE[16]. So far the attributed positive effects on improving PCE include light scattering[17-19], local field enhancement[20-22], charge transport improvement[23-25], etc. There have been many existing endeavors of doping NPs into the buffer layer (hole transport or electron transport layer)[26-28] or at the electrode/buffer interface [29-31], but the large distance between the plasmonic elements and the active material alleviate the enhancement effect in photon-to-exciton generation. Aiming for exploiting the plasmonic effect, NPs should be introduced into the active layer[32-41] or at least at the buffer/active interface.[42-46] At an early stage, there was a debate about whether bare plasmonic NPs in contact with the active materials would deteriorate the cell performance[36,37,46] or improve it.[32-34] A systematical study carried out by Salvador et al. using photoinduced absorption spectroscopy shows that there is indeed charge recombination induced near the vicinity of the metal surface, but such negative effect can be successfully inhibited by coating a thin insulating film surrounding the NPs.[35] A series of core-shell plasmonic NPs, like silica-coated Ag (Ag@SiO$_2$) nanospheres[38], silica-coated Au (Au@SiO$_2$) nanospheres[39] or nanorods (NRs)[40], et al. [41], were introduced in the active layer of OSCs, all of which displayed significant enhancement in PCEs.

In this work, within a conventional OSC device, core-shell configured plasmonic elements were introduced, specifically the Au@SiO$_2$ NRs, as they possess dual localized SP resonances, which greatly exceed nanospheres in enhancing local field. Different from literatures[40-41], we incorporate the coated Au NRs at the buffer/active interface. A previous report indicated that by introducing bare Au NRs at the buffer/active interface, the open circuit voltage ($V_{oc}$) has dropped obviously [44], thereby limiting the improvement in PCE. Here, we present that by coating the Au NRs with a thin silica film, $V_{oc}$ can be maintained very well. By optimizing the concentration of Au@SiO$_2$ NRs, we demonstrate that the plasmonic OSC outperforms the control device of 11.6% in the short circuit current ($J_{sc}$) with its external quantum efficiency (EQE) significantly enhanced over a very broad spectrum range. The fill factor (FF) of the optimized plasmonic OSC is roughly the same with that of the control cell, thus overall we achieve 11.2% improvement in PCE. The electromagnetic simulations clearly display that the field intensity surrounding Au@SiO$_2$ NRs is greatly enhanced due to the excitation of SP resonances, which should benefit the exciton generation in the active layer. In addition, we observe a strong peak in the EQE enhancement spectrum at the band edge of the active material, simulation reflects that this may come from the coupled SP resonance between neighboring NPs[43] as the spin-coating process automatically forms clustered Au@SiO$_2$ NRs on top of the PEDOT:PSS surface. Other electrical and optical characterizations further confirm that there is SP boosted exciton generation in the plasmonic OSC. Our work contributes to the development of low-cost thin-film solar cells.

## 2. Experimental Details

### 2.1. Synthesis of Au NRs

Au NRs were synthesized by the seed-mediated growth method[47]. Hydrogen tetrachloroaurate (III) hydrate (HAuCl$_4$, ≥ 49.0% Au basis), cetyltrimethylammonium bromide (CH$_3$(CH$_2$)$_{15}$N(Br)(CH$_3$)$_3$, CTAB, 99%), sodium borohydride (NaBH$_4$, 98%), and ascorbic acid were purchased from *Sigma-Aldrich*. Silver nitrate (AgNO$_3$, 99.8%) was purchased from *Sinopharm Chemical Reagent CO., Ltd*. The seed solution was synthesized first. Briefly, CTAB solution (5 mL, 0.2 M) was mixed with 0.1 mL HAuCl$_4$ (25 mM). Then, 0.6 mL of ice-cold NaBH$_4$ (0.01 M) was quickly added to the mixed solution under vigorous stirring. The seed solution was formed in brownish-yellow color 2 min later. And after 15 min stirring, it was stored at 29 °C for use. For the growth of Au NRs, 4 mL HAuCl4 (25 mM) and 0.3 mL AgNO$_3$ (16 mM) were separately added into a 100 mL aqueous solution, and then 100 mL CTAB solution (0.2 M) was added. 1.5 mL ascorbic acid (80 mM) was added to reduce Au$^{3+}$ and Au$^+$. Then, the clear growth solution was formed of colorless. Sequentially, 240 μL seed solution was added into the prepared growth solution to start the growth of Au NRs. Finally, the Au NRs were well formed after 24 h at 30 °C. The as-prepared Au NRs were centrifuged once at a speed of 9600 rpm to remove excess CTAB, and then redispersed in deionized water. By changing the amount of AgNO$_3$ solution from 0.26 mL to 1 mL, the length to diameter ratio of Au NRs can be tuned with the longitudinal SP peak shifting from 620 nm to 790 nm.

### 2.2. Preparation of Au@SiO$_2$ NRs

Preparation of silica coated Au NRs with a silica shell was carried out according to a previously published procedure.[48-49] Tetraethoxysilane (TEOS), anhydrous ethanol (99.7%), and ammonia aqueous (25%) were purchased from *Sinopharm Chemical Reagent CO., Ltd*. The pH value of 50 mL as-prepared Au NRs solution was tuned to approximately 9 by adding ammonia aqueous solution, and then 9.5 mL TEOS ethanol solution (10 mM) were added into the Au NRs solution to start the silica coating. Then, the obtained solution was agitated vigorously for 24 h using a magnetic stirrer. Au@SiO$_2$ NRs were collected by centrifugation at 6500 rpm for 15 min and washed twice with deionized water and three times with ethanol. The purified Au@SiO$_2$ NRs were redispersed into 20 mL ethanol.

### 2.3. Fabrication of OSCs and films

The control OSC has a configuration of ITO/PEDOT:PSS/PTB7:PC$_{70}$BM/LiF/Al, named as the control cell hereafter. The donor PTB7 (poly[4,8-bis[(2-ethylhexyl)oxy]benzo[1,2-b:4,5-bA]dithiophene-2,6-diyl][3-fluoro-2- [(2-ethylhexyl) carbonyl]thieno[3,4-b]-thiophenediyl]) and the acceptor PC$_{70}$BM ([6,6]-phenyl-C70-butyric-acid-methyl-ester) were purchased from *1 Material*. Chlorobenzene (CB, 99%) was purchased from *Sinopharm Chemical Reagent CO., Ltd*. PEDOT:PSS was purchased from *Heraeus*. 1, 8-Diiodooctane (DIO, 95%) was purchased from *Tokyo Chemical Industry CO., Ltd*. All chemicals were of analytical grade and used without further purification. PTB7 and PC$_{70}$BM were dissolved in CB with 3% DIO with concentrations of 10 mg/mL and 15 mg/mL, respectively. PTB7 and PC$_{70}$BM were completely dissolved after the solution mixture was stirred vigorously at 60 °C for 12 h. And the stirring need to be maintained until use. Prior to beginning the fabrication process, the ITO-coated glasses were cleaned successively with deionized water, acetone, and isopropyl alcohol for 10 min in each round, and then they were treated with ultraviolet (UV) ozone for 15 min. The PEDOT:PSS hole transporting layer was spin-coated onto the ITO-coated glass at 3000 rpm for 50 s in air. It took 10 min on a hotplate of 120 °C for the solvent evaporated and the left PEDOT:PSS film is ~30 nm thick. Next, the PTB7:PC$_{70}$BM bulk heterojunction (BHJ) active layer of ~100 nm thick was spin-coated on top of the PEDOT:PSS layer at 1000 rpm for 50 s in the glove box. To dry the active layer quickly, the samples were placed into a small vacuum chamber for 20 min. Finally, the samples were transferred into a connected thermal evaporation system (with a base pressure of < $5.0\times10^{-4}$ Pa) for the deposition of a bilayer LiF/Al(1 nm/100 nm) top electrode. During the fabrication of the hole-only devices, MoO$_3$ was also deposited in the thermal evaporation system. The multilayer films for PL measurement were made using the same methods as in OSC fabrication except the PTB7 film, for which the solution was prepared by dissolving 10 mg/mL PTB7 in CB.

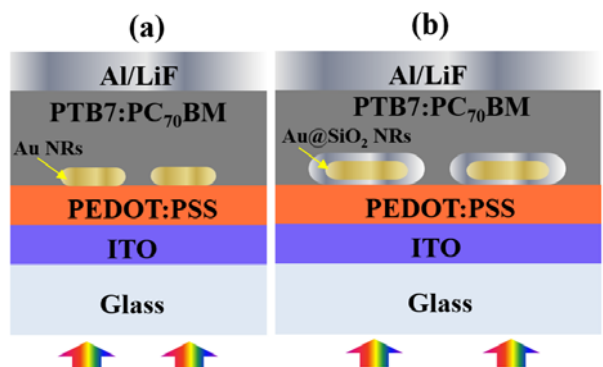

Fig. 1 Schematic diagrams of OSCs incorporating Au NRs (a) and Au@SiO$_2$ NRs (b) between the PEDOT:PSS and active layer.

The cross-sectional views of OSCs with Au NRs and Au@SiO$_2$ NRs interposed at the PEDOT:PSS/BHJ interface are shown in Fig. 1. The plasmonic interlayers were obtained by spin-coating the bare Au NRs or Au@SiO$_2$ NRs solutions in different concentrations (by the dilution method) on top of PEDOT:PSS at 1000 rpm for 40 s. In order to study the effect of Au@SiO$_2$ NRs with higher density on the performance of OSCs, the plasmonic interlayers were also made by the multiple coating approach[31] using the as-prepared Au@SiO$_2$ NRs formulation (in the concentration of 3 pM). As such, we obtained OSCs with Au@SiO$_2$ NRs of higher nominal concentrations, e.g., 9 pM and 15 pM, by repeating the spin-coating 3 and 5 times, respectively.

### 2.4. Characterization of Materials and Devices

Transmission electron microscopy (TEM) images of Au NRs and Au@SiO$_2$ NRs were taken using a transmission electron microscope at an accelerating voltage of 100 kV (JEOL JEM2100). Scanning electron microscopy (SEM) images of the surface of PEDOT:PSS incorporated with Au@SiO$_2$ NRs were measured by a thermally-assisted field emission SEM system (LEO 1530). Extinction spectra of Au NRs and Au@SiO$_2$ NRs in deionized water or ethanol solution were measured using a UV-visible spectrophotometer (Hitachi U-3900), and absorption spectra of multilayer films were also measured by Hitachi U-3900. Current density-voltage (J-V) characteristics of OSCs under AM 1.5G illumination at 100 mW/cm$^2$ and in the dark were measured using a programmable Source Meter (Keithley 2400). The illumination provided by the Sun 3000 Solar Simulator from ABET Technologies was calibrated using a standard silicon cell. J-V characteristics of hole-only devices were measured by Keithley 2400 as well. EQE of the OSCs as a function of wavelength was recorded using ZOLIX CSC1011 with a short arc xenon lamp source (Ushio UXL-553). The steady state PL spectra of ITO/PEDOT:PSS/PTB7 films with or without the presence of the Au@SiO$_2$ NRs at the PEDOT:PSS/BHJ interface were recorded by the fluorescence spectrophotometer (Tianjin Gangdong Sci.&Tech., F280).

## 3. Results and discussion

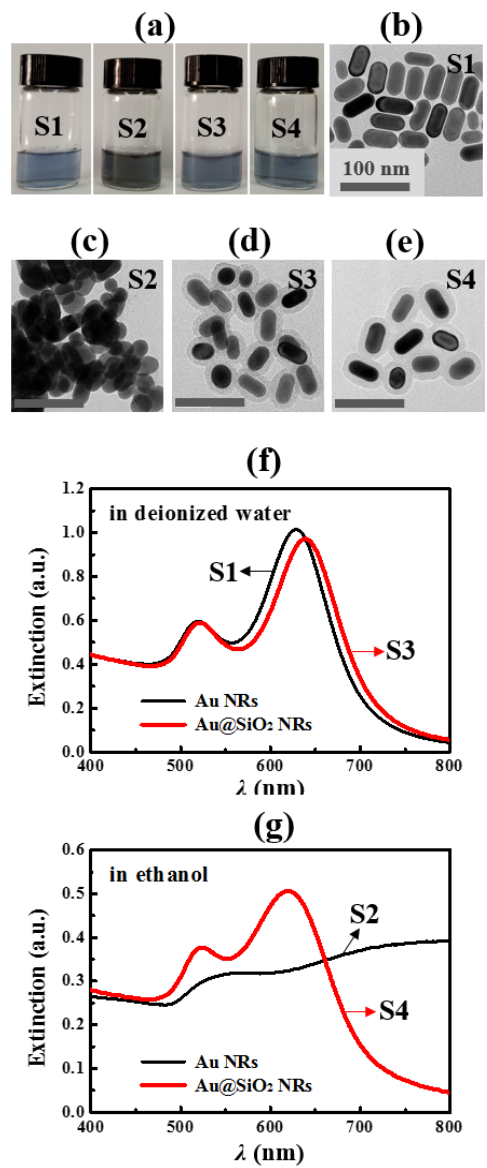

Fig. 2 Photos (a), TEM images (b-e), and extinction spectra (f-g) of four different plasmonic NP solutions. S1: Au NRs in deionized water; S2: Au NRs in ethanol; S3: Au@SiO$_2$ NRs in deionized water; S4: Au@SiO$_2$ NRs in ethanol.

The photos, TEM images, and extinction spectra of different plasmonic NP solutions are given in Fig. 2, in which S1-S4 represent the Au NRs in deionized water, the Au NRs in ethanol, the Au@SiO$_2$ NRs in deionized water, and the Au@SiO$_2$ NRs in ethanol, respectively. The prepared bare Au NRs can be well-dispersed in deionized water, yielding a clear stable solution in blue grey color as shown by S1 in Fig. 2(a). The Au NRs have an average length of about 48 nm and diameter of 24 nm [measured from Fig. 2(b)], exciting a mild transverse localized SP resonance at 519 nm and a strong longitudinal localized SP resonance at 628 nm in deionized water as shown by its extinction spectrum [thin curve, Fig. 2(f)]. The bare Au NRs is dissolved into ethanol, aiming for not deteriorating the previously-coated PEDOT:PSS film when the NRs are introduced at the buffer/active interface, but the solution quickly turns into dark grey in color [Fig. 2(a), S2]. The extinction spectrum of S2 [thin curve, Fig. 2(g)] clearly shows that the original SP resonances disappear while a very flat extinction spectra identified with no apparent peaks is produced. The TEM image of S2 [Fig. 2(c)] shows that obvious aggregation of Au NRs occurs in ethanol solution, resulting in the dark color solution and abnormal extinction spectrum. The behavior of Au NR aggregation in ethanol solution was not described in Ref. 44, although a similar Au NR synthesis method was used compared to the formulation discussed in the present work.

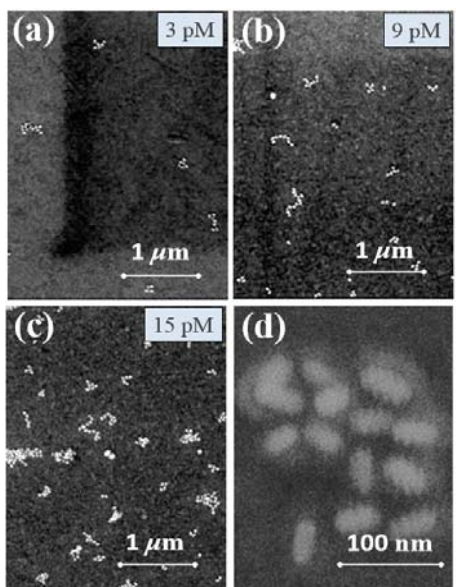

Fig. 3 (a-c) SEM images of the PEDOT:PSS surfaces incorporated with Au@SiO$_2$ NRs in different concentrations. (d) An enlarged SEM image.

Based on our results, it is found that the undesirable aggregation of Au NRs can be prevented by coating the NRs with a thin silica shell, i.e., forming Au@SiO$_2$ NRs. Fig. 2(a) reveals that Au@SiO$_2$ NRs in deionized water (S3) and in ethanol (S4) appear as clear liquid with their color similar to that of S1. TEM images of S3 [Fig. 2(d)] and S4 [Fig. 2(e)] verify that the silica-coated Au NRs can be fully dispersed in both solutions without aggregation. An interesting feature is that the silica-coated Au NRs tend to form clusters instead of appearing individually. From the TEM images, the thickness of the silica shell is ~8 nm. The extinction spectrum of Au@SiO$_2$ NRs in deionized water (S3) [thick, Fig. 2(f)] indicates that the two SP resonances are red-shifted to 522 nm and 638 nm, respectively, as a result of an increase in the refractive index of the medium surrounding the Au NRs due to the formation of the silica shell. The extinction spectrum of Au@SiO$_2$ NRs in ethanol does not have obvious change [thick curve, Fig. 2(g)].

SEM images measured for the PEDOT:PSS surfaces with plasmonic interlayers in different densities of the Au@SiO$_2$ NRs are shown in Figs. 3(a)-3(c). It is found that most of the rods are lying parallel to the PEDOT:PSS surface and the NRs are more likely to form clusters instead of appearing individually, similar to the distribution of the NPs observed in Figs. 2(d) and 2(e). Increase in the concentration yields a higher possibility of the big clusters. An enlarged SEM image, as shown in Fig. 4(d), clearly reveals the clustering effect as well as the core-shell structured Au NRs. From the SEM images shown in Figs. 3(a)-3(c), the area densities of the Au@SiO$_2$ NRs on PEDOT:PSS surfaces are estimated as ~6, ~18, and ~30 NRs per unit area of 1 μm$^2$, respectively, linearly correlated to the nominal concentrations of Au@SiO$_2$ NRs (3 pM, 9 pM, and 15 pM) in ethanol.

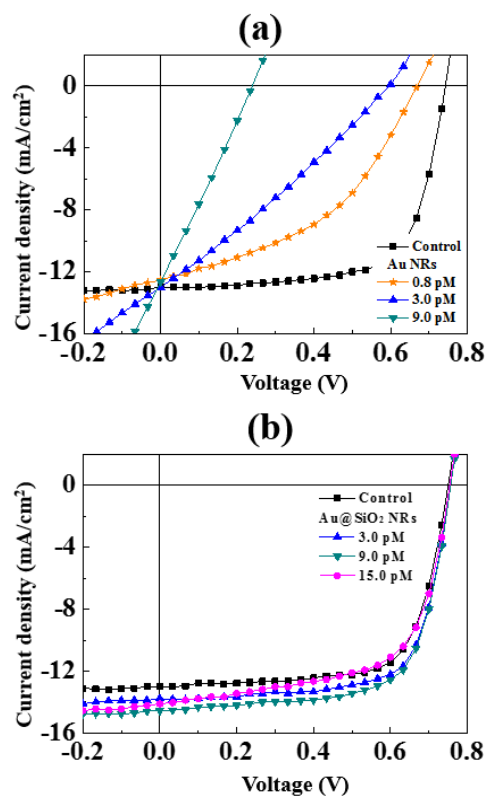

Fig. 4 *J-V* characteristics of OSCs with Au NRs (a) or Au@SiO$_2$ NRs (b) in different concentrations, under AM 1.5G illumination at 100 mW/cm$^2$.

Table 1 Device characteristics of OSCs incorporating different NRs in different concentrations.

| Device | | $V_{oc}$ (V) | $J_{sc}$ (mA/cm$^2$) | FF (%) | PCE (%) | $R_s$ (Ω·cm$^2$) | $R_{sh}$ (Ω·cm$^2$) |
|---|---|---|---|---|---|---|---|
| Control | | 0.751 | 13.01 | 70.3 | 6.87 | 5.72 | 762.84 |
| w/ Au NRs | 0.8 pM | 0.669 | 12.55 | 43.3 | 3.63 | 21.02 | 123.89 |
| | 3.0 pM | 0.597 | 13.02 | 28.1 | 2.18 | 25.09 | 53.53 |
| | 9.0 pM | 0.239 | 12.65 | 26.2 | 0.79 | 16.75 | 20.68 |
| w/ Au@SiO$_2$ NRs | 3.0 pM | 0.757 | 13.80 | 71.3 | 7.44 | 4.49 | 599.60 |
| | 9.0 pM | 0.758 | 14.52 | 69.4 | 7.64 | 4.81 | 450.00 |
| | 15.0 pM | 0.756 | 14.12 | 62.8 | 6.71 | 5.00 | 283.97 |

*J-V* characteristics measured for OSCs incorporated with bare Au NRs and Au@SiO$_2$ NRs under illumination of 100 mW/cm$^2$ are plotted in Figs. 4(a) and 4(b). The corresponding *J-V* characteristics of the OSC without the presence of NPs (i.e., the control cell) are also indicated for comparison. Performance parameters of different OSCs are summarized in Table 1. For the control cell, $V_{oc}$, $J_{sc}$, FF, and PCE are 0.751 V, 13.01 mA/cm$^2$, 70.3%, and 6.87%, respectively. The obtained PCE of the control cells distributes in a range between 6.52% and 6.87% (rectangles, Fig. 5) and the best control cell is selected for illustration in Fig. 4 and Table 1. Fig. 4(a) reveals that the OSCs with bare Au NRs perform much worse compared with the control cell, which mainly attributed to the reduced $V_{oc}$ and FF. The increase in the concentration of bare Au NRs from 0 to 9.0 pM does not bring obvious change in $J_{sc}$ but result in a dramatic drop in $V_{oc}$ from 0.743 V to 0.239 V. At a low Au NR concentration, e.g., 0.8 pM, the fabricated device still performs like a photodiode, but with an increased series resistance ($R_s$) and a decreased shunt resistance ($R_{sh}$), thereby a reduced FF. However, the characteristics of the devices made with high concentration of Au NRs, e.g., 3.0 pM and 9.0 pM, are more like a resistor because conducting channels are formed between the front and back electrodes caused by the strong aggregation of Au NRs. All these results reflect that the modification of PEDOT:PSS surface with a plasmonic interlayer made of bare Au NRs has limitation for application in OSCs.

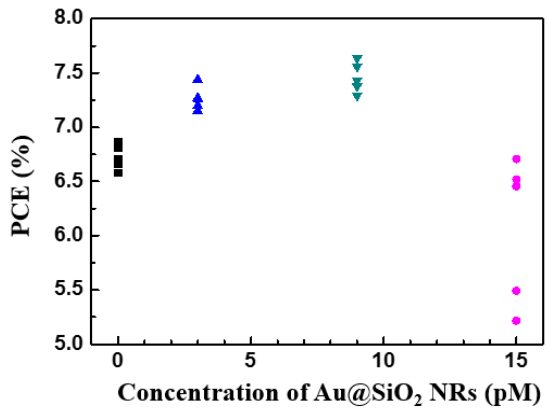

Fig. 5 Distributions of PCE of OSCs incorporated with Au@SiO$_2$ NRs in different concentrations

Instead, after coating the Au NRs with a thin silica shell, the corresponding plasmonic OSCs can considerably outperform the control cell as displayed by the distributions of their PCEs when the concentration of Au@SiO$_2$ NRs is tuned (Fig. 5). It is found that the champion cells are made with Au@SiO$_2$ NRs in 9.0 pM concentration with the highest PCE of 7.64%, 11.2% higher over that of the control cell (6.87%). The fluctuation in the PCE of the plasmonic OSCs with relatively low concentration (e.g., 3.0 pM and 9.0 pM) is similar to that of the control cell (±0.17%). But when the concentration of Au@SiO$_2$ NRs reaches to 15 pM, the fabricated OSCs have pronounced fluctuation in PCE (from 5.21% to 6.71%) as indicated by circles in Fig. 5, which might be caused by the possible stacking of NRs formed during multiple spin-coating processes. In Fig. 4(b), the displayed *J-V* curve at each concentration of Au@SiO$_2$ NRs corresponds to the cell of which the PCE is the highest among its kind as shown in Fig. 5. The first important feature in Fig. 4(b) is that $V_{oc}$ is maintained very well for all of the demonstrated plasmonic OSCs. It is also clear that the plasmonic OSCs possess a higher $J_{sc}$ compared to that of the control cell, responsible for the enhanced PCE. The highest $J_{sc}$ of 14.52 mA/cm$^2$ was obtained for the OSC incorporating 9.0 pM Au@SiO$_2$ NRs, displaying a 11.6% increase over that of the control cell. From Table 1, it is found that when the concentration of Au@SiO$_2$ NRs is low (e.g., 3.0 pM), the corresponding OSC has a slight reduction in both $R_s$ and $R_{sh}$, resulting in the FF even a bit higher than that of the control cell. However, when the concentration of Au@SiO$_2$ NRs is 15 pM, $R_{sh}$ decreases significantly, bringing ~10% drop in FF with respect to the control cell. For our champion cell, there is a small decrease in $R_s$ and an obvious decrease in $R_{sh}$, respectively, compared to the control cell, overall leading to a small reduction in FF (1.2%). But the enhancement of $J_{sc}$ overwhelms the reduction of FF so that our champion plasmonic cell can outperform the control cell by 11.2 % in PCE.

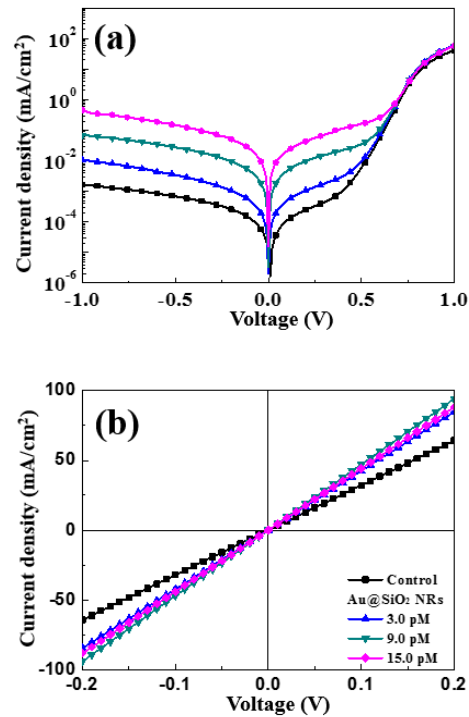

Fig. 6 J-V characteristics for (a) the OSCs with and without Au@SiO$_2$ NRs in the dark and (b) hole-only devices of ITO/PEDOT:PSS/MoO$_3$/Ag with and without Au@SiO$_2$ NRs on top of the PEDOT:PSS surface

Fig. 6(a) plots *J-V* characteristics for the OSCs with and without Au@SiO$_2$ NRs in the dark. It is seen that the current rectification ratio (forward to reverse bias current at ±1 V) decreases gradually and the leakage current increases gradually with the increase of the concentration of Au@SiO$_2$ NRs, consistent with the gradual reduction in the shunt resistance as indicated in Table 1. This must be associated with the interface defects introduced after incorporating Au@SiO$_2$ NRs on top of PEDOT:PSS. From the SEM images as shown in Figs. 3(a)-3(c), it is known that the particle density at the PEDOT:PSS/BHJ interface increases with the Au@SiO$_2$ NR concentration. The larger number of nanoparticles introduced into the OSC, the higher density of the defects at the interface, increasing the charge recombination and therefore the leakage current.

To further explain the change in series resistance between the control cell and the plasmonic OSCs, a series of hole-only devices with a configuration of ITO/PEDOT:PSS(100 nm)/MoO$_3$(5 nm)/Ag, with or without the Au@SiO$_2$ NRs presented at the PEDOT:PSS/BHJ interface, was made for comparison. The corresponding J-V characteristics are shown in Fig. 6(b). It can be clearly observed that the slopes of the J-V curves of all plasmonic hole-only devices are larger than that of the device without any NPs, indicating the devices with the addition of Au@SiO$_2$ NRs have higher conductivity. This consists with the decrease in series resistance of OSCs with Au@SiO$_2$ NRs, as listed in Table 1. The improved performance in series resistance could be attributed to the introduction of the so-called dopant levels[19] within the band gap of the PEDOT:PSS, due to the surface modification using Au@SiO$_2$ NRs, resulting in a decrease in the energy barrier at the PEDOT:PSS/BHJ interface, thereby improving hole extraction.

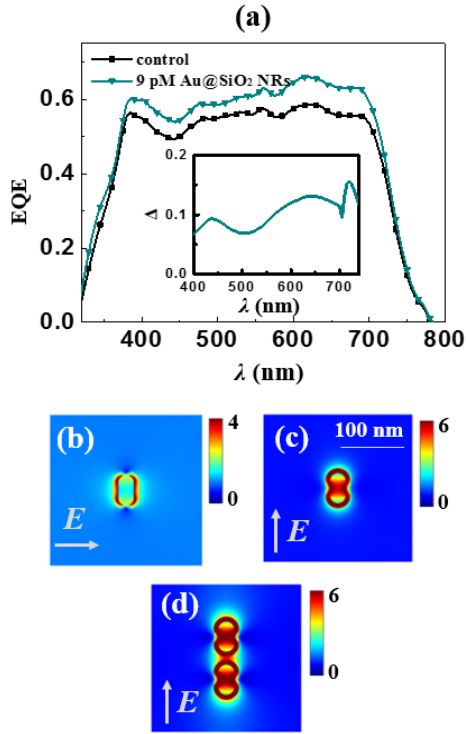

Fig. 7 (a) External quantum efficiency (EQE) versus the wavelength for the plasmonic OSC with Au@SiO$_2$ NRs (9 pM) and the control cell. The inset displays the enhancement factor of EQE for the plasmonic OSC with respect to the control cell. (b-d) Electric field distributions surrounding the Au@SiO$_2$ NRs (single or pair) at wavelengths of 560 nm, 680 nm and 720 nm, respectively.

In the following, we focus on comparing the performances of the champion plasmonic OSC (with 9.0 pM Au@SiO$_2$ NRs) and the control cell, aiming for better understanding the plasmonic induced increase in $J_{sc}$ as well as addressing the corresponding reason. First, the EQE versus the wavelength of both cells were measured and displayed in Fig. 7(a). For the device with Au@SiO$_2$ NRs (9 pM), the EQE is higher than that of the control cell over a broadband wavelength range, covering almost the whole absorbing band of PTB7:PC$_{70}$BM. The increase in EQE (indicated by Δ), defined as the difference between the EQE of the plasmonic and control cells divided by that of the control cell, is plotted in the inset of Fig. 7(a). It is observed that there is an average of 10% increase in EQE of the cell with Au@SiO$_2$ NRs (9.0 pM). $J_{sc}$ calibrated by EQE for the champion plasmonic OSC is 13.72 mA/cm$^2$, having a 10.4% increase over that of the control cell (12.42 mA/cm$^2$), consisting with the results from the J-V characteristic measurements.

Next, the enhancement of EQE can be explained with resorting to electromagnetic calculations. First, a strong enhancement in EQE within the wavelength range from 500 nm to 700 nm where the SP resonances of Au@SiO$_2$ NRs locate, is observed. By calculation, we find that the peaks of localized SP resonances of the Au@SiO$_2$ NRs are red-shifted to 560 nm and 680 nm, respectively, after introducing the NRs into the present OSC system. Electric field distributions at the two wavelengths around the Au@SiO$_2$ NRs with the displayed plane parallel to the PEDOT:PSS surface are plotted in Figs. 7(b) and 7(c), respectively. The polarizations of the incident electric field at 560 nm and 680 nm are along the short-axis and long-axis of the NR, respectively, aiming for exciting the corresponding SP resonance. In both figures, a strong enhancement in electric field localized near the vicinity of the NRs is clearly observed. We deduce that the enhanced electric field must be responsible for the improved exciton generation in the active layer. It is observed that the enhancement in the EQE spectrum over the wavelength range from 500 nm to 700 nm has some deviation from the extinction spectrum of Au@SiO$_2$ NRs, which might be owing to the clustering effect of Au@SiO$_2$ NRs and/or the change in electrical property, e.g., charge recombination, charge transport, etc., discussed above. At wavelength longer than 700 nm, there is a strong peak in the EQE enhancement spectrum (Δ) as shown in the inset of Fig. 7(a). Being aware that the Au@SiO$_2$ NRs tend to form clusters on top of the PEDOT:PSS surface, we then study a simple example with two particles linked end to end with a 5 nm separation in simulation. Under illumination at the polarization directed along the long-axis, the longitudinal SP resonances between neighboring NRs couple with each other at wavelength of 720 nm, as shown in Fig. 7(d), in agreement with the experimental results. It is seen in Fig. 7(d) that the intensity of the electric field near the vicinity of the NR pairs significantly exceeding that for a single NR [Fig. 7(c)].[15,43] For the minor peak in the EQE enhancement spectrum (Δ) at wavelength shorter than 500 nm, also reported in Refs. 50-52, so far we do not have a clear explanation for it. But from the extinction spectrum of Au@SiO$_2$ NRs [Fig. 2(a)], it is seen that the extinction of light is also quite strong at this wavelength region due to transition between electronic bands in the bulk Au. It is proposed that the inter-band transition in Au, accompanied by the generation of heat, may affect the exciton generation rate around the NPs and thereby being responsible for observed enhancement in EQE in the short wavelength range.

Fig. 8(a) shows the double logarithmic plot of photocurrent density ($J_{ph}$) as a function of the effective voltage ($V_{eff}$). Here, $J_{ph} = J_L − J_D$, where $J_L$ and $J_D$ are the current densities under illumination and in the dark, respectively; $V_{eff} = V_0 − V_a$, where $V_0$ is the built-in voltage measured at $J_{ph} = 0$, and $V_a$ is the applied bias. When the effective voltage is lower than 0.1 V, the photocurrent for both cells increases linearly as the effective voltage increases, and then tends to saturate.[53] Here, we see that the photocurrent generated at low $V_{eff}$ for the plasmonic OSC is lower than that for the control cell due to the relatively high charge recombination induced by the surface defects formed during the incorporation of Au@SiO$_2$ NRs. The photocurrent further increases with the effective voltage at higher reverse voltage ($V_{eff} > 1$ V). From the tendency of the $J_{ph} \sim V_{eff}$ curves shown in Fig. 7(e), it is deduced that the saturated photocurrent

(i.e., the photocurrent at sufficiently high effective voltage) for the plasmonic OSC is predicted much higher than that of the control cell. Under the assumption that all of the photo-generated excitons are dissociated at sufficiently high electric field, the saturated photocurrent through the external circuit then depends linearly on the maximum exciton generation rate, which is a measure of the maximum number of photons absorbed. Therefore, the elevated saturated photocurrent suggests that enhanced light absorption is produced in the active layer of the plasmonic OSC compared to the control cell.

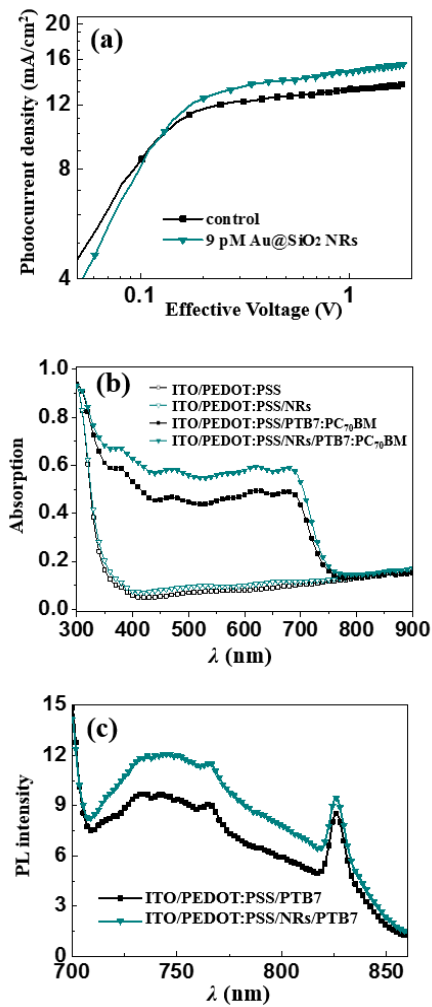

Fig. 8 (a) Photocurrent density versus the effective voltage for the plasmonic OSC and the control cell. (b) Absorption spectra of different multilayer films as indicated. (c) Steady-state photoluminescence (PL) spectra of ITO/PEDOT:PSS/PTB7 films with or without Au@SiO$_2$ NRs. All plasmonic OSCs or films incorporates Au@SiO$_2$ NRs in concentration of 9.0 pM.

In the following, different multilayer films were fabricated with their absorption spectra being characterized. The two curves with solid symbols in Fig. 8(b) show the absorption spectra of two multilayer films in the configuration of ITO/PEDOT:PSS/PTB7:PC$_{70}$BM with or without the Au@SiO$_2$ NRs (9 pM) at the PEDOT:PSS/BHJ interface. It is found that considerable enhancement in absorption over a broad wavelength range, covering almost the whole absorbing band of PTB7:PC$_{70}$BM, is identified with the presence of NRs, similar to the occurrence in EQE measurements. But it is hard to identify that the enhanced absorption is from PTB7:PC$_{70}$BM or not, because the NPs also contribute to the total absorption but they turn light into heat instead of photocurrent. Here, the absorption spectra of other two multilayer films in the configuration of ITO/PEDOT:PSS with or without the Au@SiO$_2$ NRs (9 pM) on top of the PEDOT:PSS surface [open symbols, Fig. 8(b)] were carried out for clarifying the origin of enhanced absorption. It is found that surface modification with Au@SiO$_2$ NRs also improves the absorption of the ITO/PEDOT:PSS film over a broadband wavelength range, but the absolute intensity of absorption enhancement (from the plasmonic NPs and PEDOT:PSS) is very minor. However, as seen, the absorption enhancement happens to ITO/PEDOT:PSS/PTB7:PC$_{70}$BM films with or without Au@SiO$_2$ NRs is much larger than that happens to ITO/PEDOT:PSS films. Such huge increase in absorption should not mainly come from the plasmonic components but should be attributed to the active layer. Therefore, it is confirmed that the incorporation of Au@SiO$_2$ NRs can improve the light absorption in the active layer, thereby the increase in exciton generation in OSCs.

To further identify that the incorporation of Au@SiO$_2$ NRs can bring forward the enhancement in photo-to-exciton process in the active layer, steady state PL spectra of ITO/PEDOT:PSS/PTB7 films with or without Au@SiO$_2$ NRs (9 pM) at the PEDOT:PSS/PTB7 interface were measured. Here, the PC$_{70}$BM acceptor is excluded in the PL measurement aiming for eliminating the process of non-radiative transition of excitons. As a result, the fluorescence intensity is mainly dependent on the density of photo-generated excitons in PTB7. The measured results at the excited wavelength of 680 nm are shown as an example [Fig. 8(c)], indicating that the PTB7 with adjacent Au@SiO$_2$ NRs has much higher PL intensity than that of the pure PTB7 film. The PL measurements confirm that the plasmonic NPs indeed enhance the absorption in PTB7 by exciting the localized SP resonances, thereby the increase in the exciton generation rate in PTB7.

## 4. Conclusions

The results of this work demonstrated that the decrease in the open circuit voltage of the OSCs with bare Au NRs at the PEDOT:PSS/BHJ interface is caused by the aggregation of bare Au NRs in ethanol solution. Such aggregation of Au NRs can be prevented by adopting Au@SiO$_2$ core-shell structured NRs. The use of the Au@SiO$_2$ NRs at the PEDOT:PSS/BHJ interface enables ~11% increase in $J_{sc}$ and PCE in PTB7:PC$_{70}$BM-based OSCs. It also reflects that with the presence of Au@SiO$_2$ NRs, both the shunt resistance and series resistance are reduced; overall, the fill factor of the optimized plasmonic OSC only has a negligible reduction than that of the control cell. The increase in $J_{sc}$ corresponds to considerable enhancement in EQE over a broad wavelength range, covering the whole absorbing band of PTB7:PC$_{70}$BM. Electromagnetic simulations reveal that the excitation of transverse and longitudinal SP resonances yields strong electric field near the vicinity of the coated NRs, responsible for the increase in exciton generation in the active layer over the wavelength range from 500 nm to 700 nm. Moreover, the automatically formed Au@SiO$_2$ NR clusters can generate even stronger electric field at the wavelength of ~720 nm than the bare NRs, due to the mutual coupling of longitudinal SP resonances produced by the closely packed NPs, resulting in a strong enhancement peak in EQE at the corresponding wavelength. The subtle designed electrical and optical measurements confirms that the addition of plasmonic NPs indeed benefits the absorption in the active layer, thereby improve the exciton generation in OSCs. This work contributes to a comprehensive understanding on the mechanism of plasmonic improved OSCs along with the development of thin film photovoltaics.

## Acknowledgements

This research work was financially supported by National Natural Science Foundation of China (11204205, 61475109, 61274056, 61275037, and 61405062), Outstanding Young Scholars of Shanxi Province, Hong Kong Scholar Program (XJ2013002), Doctoral Program of Higher Education Research Fund (20121402120017), and China Postdoctoral Science Foundation (2014M550152, 2013M530368, and 2014T70818). We acknowledge Qingyi Yang and Weixia Lan for SEM characterizations, and Prof. Ping Wang for the use of PL measurement equipment. Cui also acknowledge helpful discussions with Dr. Qingkun Liu and Dr. Bo Wu.